\documentclass[twocolumn,secnumarabic,amssymb, nobibnotes, aps, prl]{revtex4-2}

\usepackage[dvipsnames, svgnames, x11names]{xcolor}
\usepackage{rotating}
\usepackage{lipsum}
\usepackage{graphicx}
\usepackage{dcolumn}
\usepackage{bm}
\usepackage{color}
\usepackage{diagbox}

\usepackage{tabularx}

\usepackage{makecell}
\usepackage[none]{hyphenat}
\usepackage{lineno}
\usepackage{mathrsfs}

\definecolor{Navy}{RGB}{54,100,139}
\usepackage[colorlinks,
linkcolor=Navy,
anchorcolor=Navy,
citecolor=Navy]{hyperref}
\usepackage[version=4]{mhchem}
\usepackage{amsmath}
\usepackage{bm}

\usepackage{changes}
\definechangesauthor[name={HI}, color=orange]{HI}

\begin{document}
	\setlength {\marginparwidth} {2cm}
	
	
	\title{Tailoring magnetism of nanographenes via tip-controlled dehydrogenation
	}

	\author{Chenxiao Zhao$^{1{*},\textcolor{Navy}{\dagger}}$, Qiang Huang$^{2{*}}$, Leo{\v{s}} Valenta$^{3}$, Kristjan Eimre$^{1}$, Lin Yang$^{2}$, \\Aliaksandr V. Yakutovich$^{1}$, Wangwei Xu$^{1,4}$, Ji Ma$^{2,5}$, Xinliang Feng$^{2,5}$, Michal Jur{\'\i}{\v{c}}ek$^{3}$,\\ Roman Fasel$^{1,4}$, Pascal Ruffieux$^{1,\textcolor{Navy}{\ddagger}}$, Carlo A. Pignedoli$^{1,\textcolor{Navy}{\S}}$}
	\affiliation{$^{1}$Empa—Swiss Federal Laboratories for Materials Science and Technology, Dübendorf, Switzerland.}
	\affiliation{$^{2}$Faculty of Chemistry and Food Chemistry, and Center for Advancing Electronics Dresden, Technical University of Dresden, Dresden, Germany.}
	\affiliation{$^{3}$Department of Chemistry, University of Zurich, Winterthurerstrasse 190, 8057 Zurich, Switzerland.} 
    \affiliation{$^{4}$University of Bern, Bern, Switzerland.}
	\affiliation{$^{5}$Max Planck Institute of Microstructure Physics, Weinberg 2, 06120, Halle, Germany.}
    

	\begin{abstract}
		\textbf{
			Atomically precise graphene nanoflakes, called nanographenes, have emerged as a promising platform to realize carbon magnetism. Their ground state spin configuration can be anticipated by Ovchinnikov-Lieb rules based on the mismatch of $\pi$-electrons from two sublattices. While rational geometrical design achieves specific spin configurations, further direct control over the $\pi$-electrons offers a desirable extension for efficient spin manipulations and potential quantum device operations. To this end, we apply a site-specific dehydrogenation using a scanning tunneling microscope tip to nanographenes deposited on a Au(111) substrate, which shows the capability of precisely tailoring the underlying $\pi$-electron system and therefore efficiently manipulating their magnetism. 
            Through first-principles calculations and tight-binding mean-field-Hubbard modelling, we demonstrate that the dehydrogenation-induced Au-C bond formation along with the resulting hybridization between frontier $\pi$-orbitals and Au substrate states effectively eliminate the unpaired $\pi$-electron. 
            Our results establish an efficient technique for controlling the magnetism of nanographenes.}
	\end{abstract}
	\maketitle


	\sloppy{}
	The magnetism of nanographenes, rooting from the underlying $\pi$-electron systems, has garnered significant attention from the fields of physics, chemistry, and nanotechnology by virtue of their long coherence time, negligible magnetic anisotropy, and great scalability of carbon-based nanomaterials \cite{trauzettel2007spin, yazyev2008magnetic, yazyev2010emergence,lombardi2019quantum, song2021surface}.  For an extended period, the synthesis and characterization of magnetic nanographenes have posed significant challenges owing to their high reactivity. Recent advancements in on-surface synthesis have enabled the fabrication of a number of prototypical quasi-zero-dimensional (0D) magnetic nanographenes on Au (111) surface, like Clar’s goblet\cite{mishra2020topological}, [$n$]-triangulenes \cite{pavlivcek2017synthesis, su2019atomically,mishra2019synthesis}, [$n$]-rhombenes \cite{mishra2021large,biswas2023steering}, which have been characterized with respect to their magnetic ground states and spin excitations by means of scanning tunneling microscopy (STM) and inelastic electron tunneling spectroscopy (IETS) \cite{PhysRevLett.102.256802}. Additionally, coupled spin systems achieved through covalent connections between these 0D building blocks have also been realized\cite{mishra2020collective,hieulle2021surface,cheng2022surface,mishra2021observation}, offering promising platforms for spin-based quantum information processing.
 
    In general, the ground state magnetization of nanographenes can be anticipated using the Ovchinnikov-Lieb rule: $S$=$|N_{A}-N_{B}|/2$, where $S$ represents the total spin, and $N_A$ and $N_B$ correspond to the numbers of carbon atoms residing on the A and B sublattices of the honeycomb structure \cite{ovchinnikov1978multiplicity, lieb1989two}.  This can be readily understood based on the fact that each C atom in graphene contributes one $\pi$-electron with the spins parallel within the same sublattice and antiparallel with respect to the other sublattice. As a result, a sublattice imbalance gives rise to unpaired $\pi$-electrons and therefore a net magnetization, as in the triangulene family\cite{pavlivcek2017synthesis, su2019atomically,mishra2019synthesis}. Additionally, for nanographenes with balanced sublattices, unpaired $\pi$-electrons may also exist owing to topological frustration, as in Clar's goblet\cite{wang2009topological},  or due to sufficiently strong correlation effects, like in [$n$]-rhombenes \cite{biswas2023steering}. Given the pivotal role that $\pi$-electrons play in the magnetism of nanographenes, it is imperative to have the capability to manipulate the $\pi$-electron system in order to exert control over the magnetism.

    Here, by leveraging the atomic-scale manipulation capabilities of STM, we realized precise tailoring of the $\pi$-electron systems in Clar's goblet (hereinafter referred to as goblet) and [3]-triangulene via selective tip-induced dehydrogenation, which allows us to efficiently manipulate their ground state magnetization. 
    Utilizing first-principles calculations and tight-binding mean-field-Hubbard (TB-MFH) model, we elucidate that the dehydrogenation-induced Au-C bond leads to the realignment of frontier $\pi$-orbitals through strong hybridization with the Au substrate states. This, in turn, triggers the redistribution of related unpaired $\pi$-electrons, resulting in the removal of one unpaired $\pi$-electron with each dehydrogenation process. In terms of the final outcome, from the perspective of a TB-MFH model, this dehydrogenation is equivalent to removing the dehydrogenated carbon site. As a result, selective tip-induced dehydrogenation offers a precise engineering approach for manipulating the spin configuration of nanographenes.
	
	\begin{figure}
		\includegraphics[width=8.5cm]{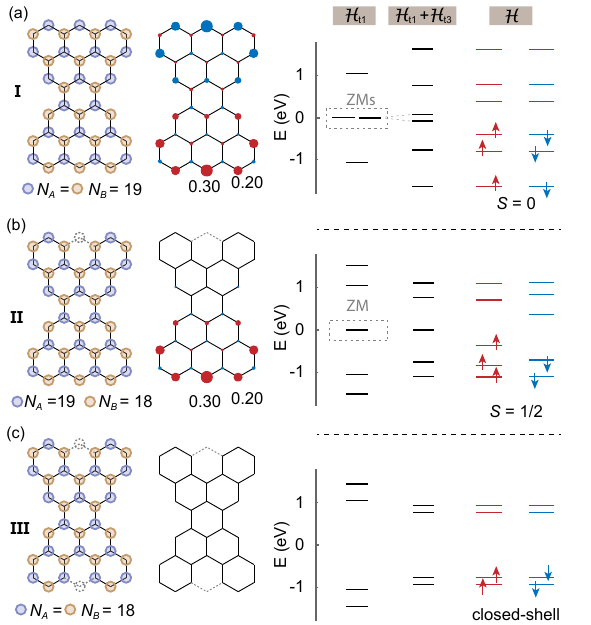}
		\caption{\label{fig1}
			Effect of removing specific carbon sites on the ground state magnetization of a goblet. (a)-(c) Sublattice-resolved schematic representations, the spin density of the ground states, and the energy spectra of a goblet (\textbf{I}), goblet with one (\textbf{II}) and two (\textbf{III}) eliminated carbon sites. Dashed circles represent the removed sites. The magnetization values of selected carbon sites are given in units of $\mu$$B$. $\mathcal{H}_{t_1}$ and $\mathcal{H}_{t_3}$ denote the kinetic terms considering the nearest-neighbor and the third-nearest-neighbor hoppings, respectively. $\mathcal{H}$ represents the TB-MFH Hamiltonian. ZM denotes zero mode.}
	\end{figure}
	
	\sloppy{}
	Fig. \textcolor{Navy}{1(a)} shows the structure of a goblet (\textbf{I}) which has balanced sublattices. The low-energy spin physics of such nanographenes can be qualitatively captured by a TB-MFH model \cite{ortiz2019exchange,mishra2020topological}:
	\begin{equation}
		\begin{aligned}
			&\mathcal{H}= \mathcal{H}_{t_1} + \mathcal{H}_{t_3} + \mathcal{H}_{U}\\
			&\mathcal{H}_{t_1}= -t_1 \sum\limits_{\left\langle i,j\right\rangle ,\sigma} {\hat{c}}^{\dagger}_{i,\sigma}{\hat{c}}_{j,\sigma}\\ &\mathcal{H}_{t_3}= -t_3 \sum\limits_{\left\langle \left\langle \left\langle i,j\right\rangle \right\rangle \right\rangle ,\sigma} {\hat{c}}^{\dagger}_{i,\sigma}{\hat{c}}_{j,\sigma}\\
			&\mathcal{H}_{U}= U\sum\limits_{i} [\sum\limits_{\sigma} \left\langle {\hat{n}}_{i,\sigma}\right\rangle {\hat{n}}_{i,\overline{\sigma}}  -  \left\langle {\hat{n}}_{i,\uparrow}\right\rangle  {\left\langle {\hat{n}}_{i,\downarrow}\right\rangle}] 
		\end{aligned}
	\end{equation}
	where $t_1=2.7$ eV and $t_3=0.27$ eV are the first- and third-nearest-neighbor hopping parameters, $U=3.2$ eV is the Coulomb repulsion between electrons occupying the same site, ${\hat{n}}_{i,\sigma}=c^{\dagger}_{i,\sigma}c_{i,\sigma}$ is the occupation number operator, and $\langle i,j \rangle$ and $\langle\langle\langle i,j \rangle\rangle\rangle$ indicate the restriction of sites $i$ and $j$ to the first- and third-nearest neighbors, respectively.  The calculated spin density of \textbf{I} shows a clear spatial and sublattice polarization associated with two unpaired $\pi$-electrons arising from topological frustration [see Fig. \textcolor{Navy}{1(a)}]. From the viewpoint of the energy spectrum, each unpaired $\pi$-electron gives rise to one zero mode (ZM) when considering only the kinetic term $t_1$. Enabling also $t_3$ leads to an energy splitting of these ZMs due to hybridization between them\cite{jacob2022theory}. Further including the Coulomb repulsion $U$ renders this hybridization energetically unfavorable, resulting in a large gap separating the created singly occupied and unoccupied spin-polarized molecular orbitals (SOMOs and SUMOs). Based on the ground state spin polarization and the energy spectrum, we can conclude that structure \textbf{I} has an $S=0$ singlet ground state which has been confirmed by the observation of singlet-triplet spin excitation in IETS measurements \cite{mishra2020topological}.  
	
	A sublattice imbalance arises in structure \textbf{II} by removing one carbon site from \textbf{I} [Fig. \textcolor{Navy}{1(b)}]. In this case, the ZM from the minor sublattice is missing, therefore no splitting is present after involving the $t_3$ channel. The TB-MFH calculation indicates an $S=1/2$ ground state with a spin-polarized orbital residing at the majority sublattice of the undefective side. Further removal of the site on the opposite terminus results in a closed-shell structure (\textbf{III}), where all the remaining $\pi$-electrons can be paired up and no spin-polarized orbital is left [see Fig. \textcolor{Navy}{1(c)}].
	
	\begin{figure}[!h]
		\includegraphics[width=8.5cm]{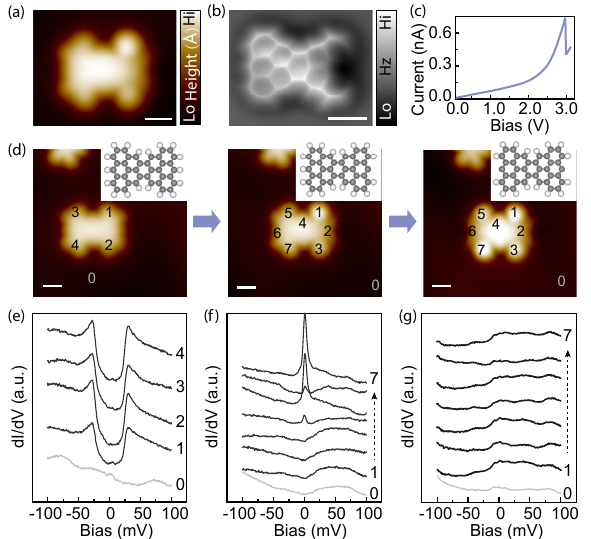}
		\caption{\label{Fig2}
			Tip-induced dehydrogenation of a goblet. (a) and (b) STM topography and nc-AFM frequency shift images of an OSD-goblet on Au(111). For STM topography: $V_{\text{bias}}$= -0.12 V, $I_{\text{set}}$= 500 pA. (c) The current discontinuity in the I-V curve indicates the breaking of the C-H bond. (d) Sequential tip-controlled dehydrogenation of the pristine goblet to the OSD-goblet and then to the BSD-goblet on Au(111) surface. Parameters: $V_{\text{bias}}=-0.05$ V, $I_{\text{set}}=500$ pA for all three images.  (e)-(g) Sequentially measured IETSs on corresponding structures shown in (d). All IETSs are taken by W tip with $V_{\text{mod}}=2$ mV (681 Hz). $I{_\text{set}}=500$ pA for (e) and $I_{\text{set}}=600$ pA for (f) and (g). All data are taken at 4.5K. Scale bars: 0.5 nm. }
	\end{figure}
	
	The above analysis raised a pathway towards manipulating spin configurations of nanographenes by intentionally inducing carbon vacancies at specific sites. To date, two processes observed in STM experiments have shown the ability to tailor the $\pi$-electron system of nanographenes, effectively creating carbon vacancies: hydrogenation and pinning the unpaired $\pi$-electron to the elbow position of the herringbone reconstruction of Au(111) \cite{mishra2020topological,li2019single, wang2022aza}. However, the former does not allow the engineering of a specific spin site, and the latter has stringent geometrical restrictions. Up to date, a precisely controllable method is still missing. Meanwhile, tip-induced dehydrogenation has been reported to be able to create significant geometrical distortion in different molecules \cite{komeda2004local,zhao2005controlling,bocquet06} and to affect the end states of graphene nanoribbons \cite{van2013suppression, talirz2013termini}. Considering its potential to change the hybridization of the corresponding carbon site, we reasoned it would be worthwhile to explore its influence on spin configurations of nanographenes. 
	
	Fig. \textcolor{Navy}{2(a)} and \textcolor{Navy}{2(b)} display the structural characterizations of a one-side dehydrogenated (OSD) goblet on Au(111) surface. As evidenced by the non-contact atomic force microscopy (nc-AFM) images in Fig. \textcolor{Navy}{2(b)}, the dehydrogenated side of the molecule bends towards the substrate, which is featured by a ``big eyes" motif in the corresponding STM image. The I-V curve shown in Fig. \textcolor{Navy}{2(c)} indicates the typical signature of C-H bond breaking, manifested by a sudden drop in the current with increasing the bias voltage (details of dehydrogenation in Methods\cite{SM,turco2022direct,ribar2019gram,holt2019short,yakutovich2021aiidalab,pizzi2016aiida,hutter2014cp2k,vandevondele2007gaussian,goedecker1996separable,perdew1996generalized,grimme2010consistent,hanke2013structure,hapala2014mechanism,kumar2016bader}). 
    By successively dehydrogenating both sides of a goblet, OSD and both-sides dehydrogenated (BSD) goblets can be obtained as shown in Fig. \textcolor{Navy}{2(d)}. The IETS measurements were performed on the molecule at different stages of the sequential dehydrogenation. For the pristine goblet, symmetric steps at $\sim \pm$ 23 meV are observed [Fig. \textcolor{Navy}{2(e)}], which has previously been rationalized as the excitation from the singlet ground state to the triplet excited state\cite{mishra2020topological}. 
	Notably, for the OSD-goblet, the IETS spectra taken on the dehydrogenated side exhibit a featureless profile compared to the background (grey curve), while spectra obtained from the unmodified side exhibit a sharp zero bias peak (ZBP) without spin excitation steps  [Fig. \textcolor{Navy}{2(f)}], similar to those observed in the case of the one-side hydrogenated goblet and the elbow-passivated goblet reported previously\cite{mishra2020topological}, proving the $S=1/2$ nature of the OSD-goblet on Au(111) surface. 
	Interestingly, this result aligns with the TB-MFH calculations for structure \textbf{II} [Fig. \textcolor{Navy}{1(b)}], where the ground state exhibits an $S=1/2$ configuration, and the spin-polarization predominantly extend at the unmodified side. This indicates that the effective role of tip-induced dehydrogenation is the removal of the dehydrogenated carbon site. As expected, further dehydrogenating the mirror carbon site erases the remaining unpaired $\pi$-electron, and therefore results in a featureless IETS spectrum [see Fig. \textcolor{Navy}{2(g)}].

	\begin{figure}[htb]
		\includegraphics[width=8.5cm]{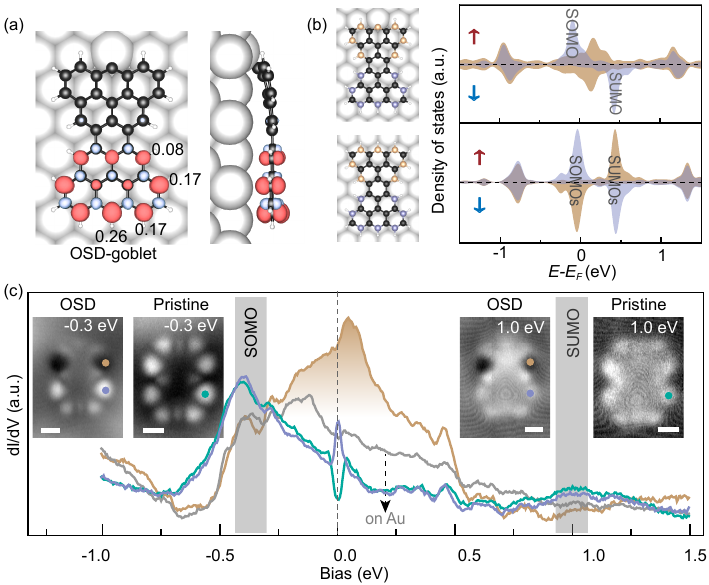}
		\caption{\label{fig3}
		(a) Optimized structure of OSD-goblet on a Au(111) slab. DFT-calculated ground state spin density is also shown. Only part of the Au atoms are shown here (see Fig. \textcolor{Navy}{S1} for the complete model\cite{SM}).   (b) Spin-resolved PDOS of the carbon atoms highlighted by corresponding colors for both OSD-goblet (upper panel) and pristine goblet (lower panel) on Au(111). (c)  Experimental dI/dV spectra [$I_{\text{set}}=600$ pA, $V_{\text{mod}}=10$ mV (681 Hz)] and dI/dV maps [($I_{\text{set}}=500$ pA, $V_{\text{mod}}=10$ mV (681 Hz)] of an OSD and a pristine goblet. The positions for taking these spectra are labeled via color-coded dots. Scale bars: 0.5 nm.}
	\end{figure}
	
	To elucidate the mechanism of the tip-induced dehydrogenation and establish a direct connection with the removal of the dehydrogenated carbon sites illustrated in Figure \textcolor{Navy}{1}, we employed density functional theory (DFT) calculations as the initial step. The optimized structure of the OSD-goblet on a Au(111) slab is illustrated in Fig. \textcolor{Navy}{3(a)}, where the dehydrogenated carbon binds to a Au atom from the substrate with a bond length of 2.11 \AA. This results in a geometrical bending of the molecule towards the substrate, in line with the experimentally observed nc-AFM image [Fig. \textcolor{Navy}{2(c)}]. The simulated nc-AFM image based on this structure fits the experimental result very well and hence confirms the validity of this structural model (see Fig. \textcolor{Navy}{S2} \cite {SM}). 
    Based on the optimized structure, the ground state spin density of an OSD-goblet on Au(111) is calculated. As shown in Fig. \textcolor{Navy}{3(a)}, the spin barely remains on the dehydrogenated side of the goblet, while it is hardly affected on the other side compared with the pristine case [see Fig. \textcolor{Navy}{S3} \cite{SM}]. This can be understood through inspection of the spin-resolved projected density of states (PDOS) shown in Fig. \textcolor{Navy}{3(b)}.  The frontier orbitals with opposite spin polarizations on the dehydrogenated side (upper panel, light-brown states) realign and pair up in energy above the Fermi level ($E_F$), while those on the unmodified side (upper panel, light purple states) are barely affected, akin to the pristine case [Fig. \textcolor{Navy}{3(b)} lower panel].  
    As a result, the original unpaired $\pi$-electron at the dehydrogenated side loses its spin polarization and is essentially transferred into the Au substrate, in accordance with the observed spin passivation. Experimentally, the realignment of the frontier orbitals at the dehydrogenated side is confirmed by dI/dV spectra [Fig. \textcolor{Navy}{3(c)}], where the original SOMO and SUMO peaks are replaced by a broad peak above the $E_F$. The spectrum taken on the unmodified side is similar to that of the pristine goblet except for the spin-related inelastic features near the $E_F$. The $dI/dV$ maps of the SOMO and SUMO for both OSD- and pristine goblets further confirm this local change of the frontier orbitals.
	\begin{figure}[h]
		\includegraphics[width=8.5cm]{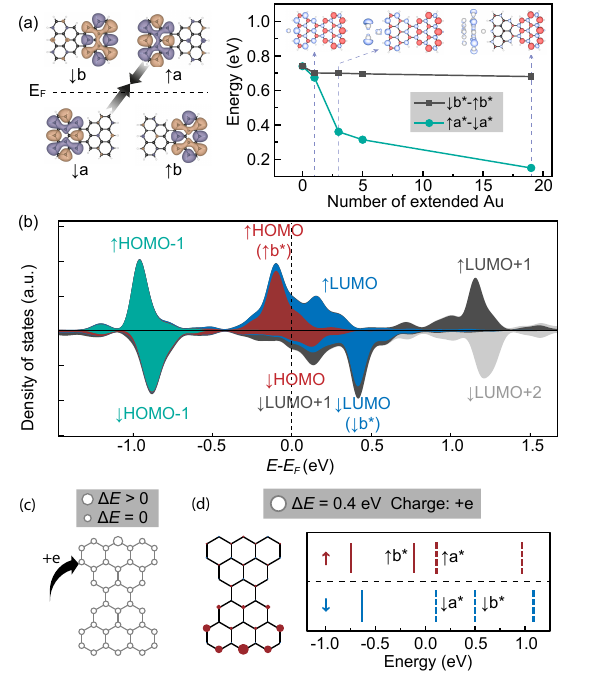}
		\caption{\label{fig4}
			  (a) DFT-calculated spin-resolved frontier $\pi$-orbitals of a freestanding pristine goblet ($\downarrow$a, $\downarrow$b, $\uparrow$b, and $\uparrow$a), along with the variations in their energy differences after one-side dehydrogenating and hybridizing with different numbers of Au adatoms (1Au-, 3Au-, 5Au-, and 19Au-bonded OSD-goblets). The corresponding orbitals after hybridizing with Au are denoted as a$^{\ast}$ and b$^{\ast}$. The ground state spin densities in 1Au-, 3Au-, and 19Au-cases are shown in the insets.  (b) Spin-resolved frontier orbitals of the OSD-goblet on Au(111). The orbitals are labeled based on the result of a freestanding OSD-goblet by removing the Au(111) slab. HOMO/LUMO: the highest (lowest) occupied (unoccupied) molecular orbital. (c) Schematic illustration of simulating an OSD-goblet on Au(111) by using a modified TB-MFH model: introducing a variation in on-site energy ($\Delta E$) at the simulated dehydrogenated carbon site and providing a global positive charge. (d) Spin density and spin-resolved energy spectrum calculated by the modified TB-MFH model.}
	\end{figure}
	
	To unveil the underlying mechanism driving this orbital realignment and gain deeper insights into the dehydrogenative spin passivation, we concentrated on the frontier $\pi$-orbitals of the pristine goblet and tracked their evolution after one-side dehydrogenating and hybridizing with varying numbers of Au adatoms in the freestanding case.
    As depicted in Fig. \textcolor{Navy}{4(a)},  the energy gap between the frontier $\pi$-orbitals on the dehydrogenated side ($\downarrow$a$^\ast$ and $\uparrow$a$^\ast$) reduces progressively upon bonding with an increasing number of Au adatoms, owing to their pronounced hybridization with the Au adatoms [details in Fig. \textcolor{Navy}{S4}]. In contrast, the gap between frontier $\pi$-orbitals at the unmodified side ($\uparrow$b$^\ast$ and $\downarrow$b$^\ast$) shows minimal variation, due to their minor hybridization with Au.
    Meanwhile, the spin polarization at the dehydrogenated side of the goblet exhibits strong leakage towards the Au adatoms, lowering the energy of the system by diminishing the overlap with the radical at the unmodified side. In the limit of a large quantity of Au atoms, namely the Au substrate case, these effects would lead to the pairing of $\downarrow$a$^\ast$ and $\uparrow$a$^\ast$ in energy above the Fermi level ($E_F$), as confirmed by the spin-resolved frontier orbitals of the OSD-goblet on Au(111) [Fig. \textcolor{Navy}{4(b)}] (see details in Fig. \textcolor{Navy}{S5}\cite{SM}). Consequently, no singly occupied spin-polarized orbital remains on the dehydrogenated side, and the formerly unpaired $\pi$-electron on this side transfers into the Au substrate.  
   
    To validate this orbital-pairing and charge-redistribution picture, we switched back to the TB-MFH model but employed a positive charge transfer +e to the molecule, instead of directly removing the dehydrogenated carbon sites [see Fig. \textcolor{Navy}{4(c)}]. To differentiate between the dehydrogenated and unmodified sides of a goblet in the model, we introduced a positive on-site energy $\Delta E= 0.4 eV$ at one carbon site to simulate the effect of the dehydrogenation-induced Au-C polar covalent bond (details in Fig. \textcolor{Navy}{S6} \cite{SM}). This $\Delta E$ also ensures that the removed $\pi$-electron originates from the simulated dehydrogenated side of the goblet.  
    The calculated ground state spin density and the spin-resolved frontier $\pi$-orbitals by this modified TB-MFH model exhibit a quantitative agreement with the DFT results [compare Fig. \textcolor{Navy}{4(d)} with Fig. \textcolor{Navy}{3(a)} and \textcolor{Navy}{4(b)}].  Notably, the orbital realignment and ground state spin density are insensitive to the value of $\Delta E$ (Fig. \textcolor{Navy}{S6}). Likewise, using +2e charge, we reproduced the closed-shell nature of the BSD-goblet on Au (see Fig. \textcolor{Navy}{S7}\cite{SM}).  
    These findings validate that the hybridization facilitated by the dehydrogenation-induced Au-C bond pairs the two spin-polarized orbitals on the dehydrogenated side and effectively eliminates the unpaired $\pi$-electron on this side. This outcome aligns with the effect of removing the dehydrogenated carbon site as depicted in Fig. \textcolor{Navy}{1}, enabling precise tailoring of the magnetism of goblets.

	\begin{figure}[ht]
		\includegraphics[width=8.5cm]{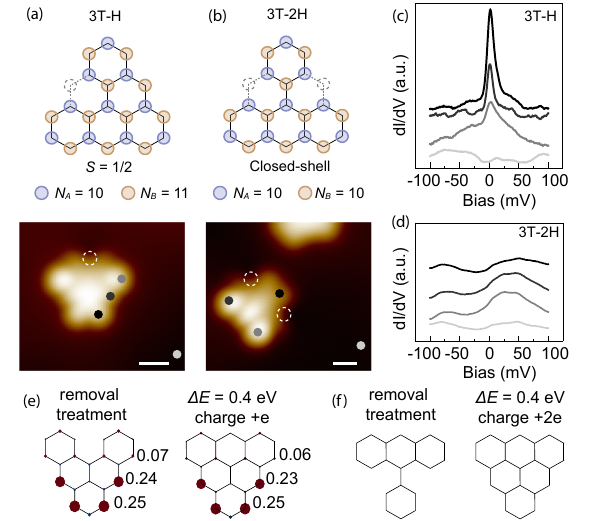}
		\caption{\label{fig5}
			Dehydrogenative tailoring of $\pi$-electron system in [3]-triangulene. (a) and (b) Schematics and the corresponding topographic images for different dehydrogenative structures: 3T-H (a), 3T-2H (b). (c) and (d) IETSs taken on 3T-H and 3T-2H, respectively. The positions of the spectra are marked by colored-coded dots. $I_{\text{set}}=600$ pA, $V_{\text{mod}}=2$ mV (681 Hz). (e) and (f) TB-MFH calculated ground state spin densities for 3T-H and 3T-2H, respectively. Both direct removal and modified treatments are used. Scale bars: 0.5 nm.}
	\end{figure}
	
	The versatility of this dehydrogenative tailoring approach has been further demonstrated through its application to [3]-triangulene (3T). Pristine 3T has an imbalanced sublattice ($N_A=10$ and $N_B=12$) and therefore a ground state with $S=1$, featuring a faint zero-energy peak that can be interpreted as an underscreened Kondo effect \cite{sasaki2000Kondo,roch2009observation,turco2022direct}. By selectively dehydrogenating specific carbon sites of 3T, we obtained structures 3T-H and 3T-2H, respectively (see \textcolor{Navy}{5(a)} and \textcolor{Navy}{5(b)}). Their ground state magnetization can be anticipated by the Ovchinnikov-Lieb rules: $S=1/2$ for 3T-H and closed-shell for 3T-2H, which are consistent with the TB-MFH results by using either direct removal or the modified treatments. These predictions are confirmed by the IETS taken on respective structures, as shown in Fig. \textcolor{Navy}{5(c)} and \textcolor{Navy}{5(d)}. Remarkably, the intensities of the Kondo resonance peaks observed at different positions on 3T-H are consistent with the spin density calculated using the TB-MFH model, as depicted in Fig. \textcolor{Navy}{5(e)}.
	
	In conclusion, our study has demonstrated the feasibility of tailoring $\pi$-electron systems and manipulating the magnetization of nanographenes through tip-induced dehydrogenation. 
    The formation of the Au-C bond after dehydrogenation and the resulting hybridization contribute to the effective tailoring, which, in terms of the final result, is equivalent to directly removing the dehydrogenated carbon site in the TB-MFH level analysis. 
    This approach is applicable for various magnetic nanographenes that have spatially separated spin-polarized $\pi$-orbitals, offering an efficient approach to selectively switch off specific spin sites in nanographene-based spin systems and hence manipulate their ground state magnetization.

	This work was supported by the Swiss National Science Foundation (grant no. 200020-182015, CRSII5$\_$205987, M.J.$/$PP00P2$\_$170534 and PP00P2$\_$198900), the NCCR MARVEL funded by the Swiss National Science Foundation (grant no. 51NF40-182892), the EU Horizon 2020 research and innovation program-Marie Skłodowska-Curie grant no. 813036 and Graphene Flagship Core 3 (grant no. 881603), ERC Consolidator grant (T2DCP, grant no. 819698), and ERC Starting grant (M.J.$/$INSPIRAL, grant no. 716139). This work was supported by a grant from the Swiss National Supercomputing Centre (CSCS) under project ID s1141.  We acknowledge PRACE for awarding access to the Fenix Infrastructure resources at CSCS, which are partially funded from the European Union’s Horizon 2020 research and innovation programme through the ICEI project under the grant agreement No. 800858. We also greatly appreciate financial support from the Werner Siemens Foundation (CarboQuant). For the purpose of Open Access (which is required by our funding agencies), the authors have applied a CC BY public copyright license to any Author Accepted Manuscript version arising from this submission.
 \\
 \\
 $^*$ equal contribution \\
\email[{$\textcolor{Navy}{^{\dagger}}$}Corresponding author:  {chenxiao.zhao@empa.ch}\\
\email[{$\textcolor{Navy}{^{\ddagger}}$}Corresponding author: {pascal.ruffieux@empa.ch}\\
\email[{$\textcolor{Navy}{^{\S}}$}Corresponding author:  {carlo.pignedoli@empa.ch}
	\bibliography{Tailoring}
	
\end{document}